\documentclass[conference]{IEEEtran}
\usepackage{epsfig}
\usepackage{times}
\usepackage{float}
\usepackage{afterpage}
\usepackage{amsmath}
\usepackage{amstext}
\usepackage{amssymb,bm}
\usepackage{latexsym}
\usepackage{color}
\usepackage{graphicx}
\usepackage{amsmath}
\usepackage{amsthm}
\usepackage{graphicx}
\usepackage[center]{caption}
\usepackage{pstricks}
\usepackage{subfigure}
\usepackage{booktabs}
\usepackage{multicol}
\usepackage{multirow}
\usepackage{lipsum}
\usepackage{dblfloatfix}
\usepackage{tabularx}
\usepackage{pbox}
\usepackage{rotating}

\newtheorem{defn}{Definition}
\newtheorem{thm}{Theorem}[section]
\newtheorem{cor}[thm]{Corollary}
\newtheorem{prop}{Proposition}
\newtheorem{lem}[thm]{Lemma}
\newtheorem{conj}[thm]{Conjecture}
\newtheorem{constr}[thm]{Construction}
\newtheorem{note}{Remark}

\newcommand{\bit}{\begin{itemize}}
\newcommand{\eit}{\end{itemize}}
\newcommand{\bcor}{\begin{cor}}
\newcommand{\ecor}{\end{cor}}
\newcommand{\beq}{\begin{equation}}
\newcommand{\eeq}{\end{equation}}
\newcommand{\beqn}{\begin{equation*}}
\newcommand{\eeqn}{\end{equation*}}
\newcommand{\bea}{\begin{eqnarray}}
\newcommand{\eea}{\end{eqnarray}}
\newcommand{\bean}{\begin{eqnarray*}}
\newcommand{\eean}{\end{eqnarray*}}
\newcommand{\ben}{\begin{enumerate}}
\newcommand{\een}{\end{enumerate}}
\newcommand{\bc}{\begin{center}}
\newcommand{\ec}{\end{center}}
\newcommand{\bdefn}{\begin{defn}}
\newcommand{\edefn}{\end{defn}}
\newcommand{\bnote}{\begin{note}}
\newcommand{\enote}{\end{note}}
\newcommand{\bprop}{\begin{prop}}
\newcommand{\eprop}{\end{prop}}
\newcommand{\blem}{\begin{lem}}
\newcommand{\elem}{\end{lem}}
\newcommand{\bthm}{\begin{thm}}
\newcommand{\ethm}{\end{thm}}
\newcommand{\bconj}{\begin{conj}}
\newcommand{\econj}{\end{conj}}
\newcommand{\bconstr}{\begin{constr}}
\newcommand{\econstr}{\end{constr}}
\newcommand{\bpf}{\begin{proof}}
\newcommand{\epf}{\end{proof}}

\begin{document}

\title{On (Secure) Information flow for Multiple-Unicast Sessions: Analysis with Butterfly Network }
\author{
\IEEEauthorblockN{Gaurav Kumar Agarwal, Martina Cardone, Christina Fragouli }
Department of Electrical Engineering \\
University of California Los Angeles, Los Angeles, CA 90095, USA\\
Email: \{gauravagarwal, martina.cardone, christina.fragouli\}@ucla.edu
\thanks{The work of the authors was partially funded by NSF under
award 1321120. 
G. K. Agarwal is also supported by the Guru Krupa Fellowship.}}
\IEEEoverridecommandlockouts

\maketitle

\begin{abstract}
This paper considers a class of wireline networks, derived from the well-known butterfly network, over which two independent unicast sessions take place simultaneously. 
The main objectives are to understand when network coding type of operations are beneficial with and without security considerations and
to derive the ultimate gains that cooperation among sources and sinks can bring.
Towards these goals, the capacity region of the butterfly network with arbitrary edge capacities is first derived. 
It is then shown that no rate can be guaranteed over this network under security considerations, when an eavesdropper wiretaps any of the links.
Three variants of the butterfly network, such as the case of co-located sources, are analyzed as well and their secure and non-secure capacity regions are characterized.
By using the butterfly network and its variants as building blocks, these results can be used to design high-throughput achieving transmission schemes for general multiple-unicast networks.
\end{abstract}

\section{Introduction}
The focus of this work is on a class of wireline networks, derived from the famous buttefly network, over which two independent unicast sessions take place simultaneously. 
Our goal is two-fold:
(i) we seek to understand when network coding type of operations are beneficial with and without security considerations and
(ii) we aim to find the ultimate gains that cooperation among sources and sinks can bring.

The characterization of the capacity of multiple-unicast wireline networks is a long-standing open problem, even for the two-unicast network. 
In particular, in~\cite{KamathISIT2014} the authors proved that solving the two-unicast problem (for general rate pairs) is as hard as solving the $k$-unicast problem, with $k \geq 3$.
For a general two-unicast network with integer edge capacities, the authors in~\cite{WangIT2010} derived necessary and sufficient conditions to achieve the point $(1,1)$. 
However, the assumption of integer edge capacities is crucial and the result does not appear to easily generalize to obtain the conditions for achieving other points, such as $(2,2)$. 
Different from muticast networks for which linear network coding suffices for capacity characterization~\cite{LiIT2003}, it is proved to be not sufficient for the case of multiple-unicast networks~\cite{dougherty2005insufficiency}, even for the two-unicast problem~\cite{KamathISIT2014}. 
This fact led to the design of several suboptimal transmission strategies. 
For example, in~\cite{TraskovISIT2006} the authors designed achievable schemes for general networks by using as a building block the famous butterfly network with uniform edge capacities, for which XORing based operations are optimal.

Since the problem of characterizing the capacity of multiple-unicast networks is open, to the best of our knowledge, the case of secure communication has not been analyzed.
For multicast traffic, in~\cite{CaiIT2011} the authors considered uniform edge capacities and showed that the cut-set bound is tight, when a passive eavesdropper has access to any $k$ channels. 
Recently, in~\cite{CuiITW2010} a more general case was considered  where edges are of arbitrary capacities; however, for this scenario the cut-set bound is not tight and hence the problem remains open even in the single unicast case. 
Thus, it is not surprising that no work considered the case of security for multiple-unicast scenarios.

In this paper, we analyze the celebrated butterfly network in Fig.~\ref{fig:butt1}, where the edge capacities are arbitrary. 
We first characterize its capacity region without security constraints, by designing a scheme that achieves the generalized network sharing outer bound derived in~\cite{kamath2011generalized}.
We then prove that secure communication is not possible over this network, when a pas- sive eavesdropper wiretaps any of the links.
We finally derive secure and non-secure capacity results for other three two-unicast networks derived from the butterfly network, namely: (i) the case of co-located sources in Fig.~\ref{fig:CS}; (ii) the case of co-located sinks in Fig.~\ref{fig:CD}; (iii) a modified version of the butterfly network in Fig.~\ref{fig:butt2}.
On the one hand, these results, by using the butterfly network and its variants as building blocks, can be used to design high-throughput achieving transmission schemes for general multiple-unicast networks with and without security considerations.
On the other hand, the results here presented provide network examples for which coding across sessions is not necessary in absence of security, but it becomes of fundamental importance under security constraints.
This observation is in line with our previous work in~\cite{agarwal2016secure} where we proved that network coding type of operations (which are not beneficial in absence of security) are crucial for characterizing the secret capacity region of three two-unicast networks with erasure channels.
Finally, the capacity results here derived shed light on the ultimate gains that can be achieved by allowing cooperation among the two sources or the two destinations.

In Section~\ref{sec:sys_mod} we define our setup. 
In Section~\ref{sec:MainResult} we derive our main result, namely we characterize the capacity for the four networks in Table~\ref{table:MessaDescr} with and without security considerations.
Finally, in Section~\ref{sec:summary} we draw conclusions and we briefly discuss how the results presented in this work can be used to design high-throughput achieving transmission schemes for a general multiple-unicast network.  

\section{Setup}
\label{sec:sys_mod}
A wireline network is represented by a directed acyclic graph $\mathcal{G} = (\mathcal{V}, \mathcal{E})$, where $\mathcal{V}$ is the vertex (node) set and $\mathcal{E}$ is the set of the directed edges. 
Each edge $ e \in \mathcal{E}$ represents a noiseless orthogonal channel of capacity $\mathsf{C}_e$. 
If an edge $e \in \mathcal{E}$ connects a node $i$ to a node $j$, we refer to node $i$ as the tail and to node $j$ as the head of the edge $e$. 
For each node $v \in \mathcal{V}$, we define $\mathcal{I}(v)$ as the set of all incoming edges of node $v$ and $\mathcal{O}(v)$ as the set of all outgoing edges of node $v$. 
	
For a two-unicast system, there are two source nodes $\mathsf{S}_1$ and $\mathsf{S}_2$ and two sink (destination) nodes $\mathsf{D}_1$ and $\mathsf{D}_2$. 
These source and destination nodes can be co-located, i.e., the two sources or/and the two destinations can be gathered together in a single node.
Each source has an independent message that has to be communicated to the corresponding destination. 
We are interested in the rates at which these messages can be reliably communicated with and without security constraints. 

Source $\mathsf{S}_i, i \in [1:2]$ has a message $W_i$ that has to be reliably decoded at node $\mathsf{D}_i$. 
The messages $W_1$ and $W_2$ are independent, uniformly drawn from a finite alphabet set and are of $q$-ary entropy rates $R_1$ and $R_2$, respectively.
Each channel is a discrete noiseless channel accepting alphabets over $\mathbb{F}_q$. 
The symbol transmitted (respectively, received) over $n$ channel uses on edge $e \in \mathcal{E}$ is denoted as $X^n_e$ (respectively, $Y^n_e$).        Clearly, since channels are noiseless, $Y_{ei} = X_{ei}, \forall i \in [1:n]$. 

\begin{defn}
\label{def:Reliability}
A rate pair $(R_1, R_2)$ is said to be achievable if there exist a block length $n$, 
a set of encoding functions $f_{e}, \  \forall e \in \mathcal{E}$, such that
	\begin{align*}
	X_{e}^n = 
	\left \{
	\begin{array}{ll}
	f_{e} \left( W_1, W_2 \right) & \text{if} \ \text{tail}(e) = \left \{\mathsf{S}_1, \mathsf{S}_2 \right \}
	\\
	f_{e} \left(W_1  \right) & \text{if} \ \text{tail}(e) = \left \{\mathsf{S}_1\right \}
	\\
	f_{e} \left(W_2  \right) & \text{if} \ \text{tail}(e) = \left \{ \mathsf{S}_2\right \}
	\\
	f_{e} \left( \{Y^{n}_\ell : \ell \in \mathcal{I}(\text{tail}(e))\}  \right) & \text{otherwise}
	\end{array}
	\right.,
	\end{align*}
	and a set of two decoding functions $\phi_j$ for $j \in [1:2]$,
such that destination $\mathsf{D}_j$ can decode $W_j$ with high probability, i.e., $\Pr \left( \phi_j \left( \left \{Y_\ell^n: \ell \in \mathcal{I}\left( \mathsf{D}_j\right) \right \}\right)\neq W_j\right) < n\epsilon_n$, $\forall \epsilon_n > 0 $.
	\end{defn}

We are also interested in finding the rate pairs at which the two messages $W_1$ and $W_2$ can be communicated securely. 
In particular, we assume that a passive eavesdropper wiretaps one channel, which one exactly is not known.
This assumption is equivalent to have one eavesdropper on every link, but these eavesdroppers do not cooperate among themselves.
We let $Z_{e}^n, e \in \mathcal{E}$ be the symbol received by the eavesdropper on edge $e$ over $n$ channel uses. 
Clearly, $X_{ei} = Y_{ei} = Z_{ei}, \forall i \in [1:n]$.
We also assume that for $j \in [1:2], \mathsf{S}_j$ has an independent and infinite source of randomness $\Theta_j$.          	


\begin{defn}
	A rate pair $(R_1, R_2)$ is said to be securely achievable if there exist a block length $n$, a set of encoding functions  $f_e, \  \forall e \in \mathcal{E}$ such that
	\begin{align*}
	X^n_e \!=\! 
	\left \{
	\begin{array}{ll}
	f_e \left( W_1, W_2, \Theta_1, \Theta_2 \right) & \text{if} \ \text{tail}(e) \!=\! \left \{\mathsf{S}_1, \mathsf{S}_2 \right \}
	\\
	f_e \left(W_1, \Theta_1  \right) & \text{if} \ \text{tail}(e) \!=\! \left \{\mathsf{S}_1 \right \}
	\\
	f_e \left(W_2, \Theta_2  \right) & \text{if} \ \text{tail}(e) \!=\! \left \{\mathsf{S}_2 \right \}
	\\
	f_e \left( \{Y^n_\ell : \ell \in \mathcal{I}(tail(e))\}  \right) & \text{otherwise}
	\end{array}
	\right.,
	\end{align*}
	and a set of two decoding functions $\phi_j$ for $j \in [1:2]$, such that destination $\mathsf{D}_j$ can reliably decode the message $W_j$ (see Definition~\ref{def:Reliability})
	and such that $\forall e \in \mathcal{E}$ and $\forall \epsilon_n > 0$
	$I \left(W_1, W_2 ; Z^n_e\right) < \epsilon_n$ (strong secrecy requirement).
\end{defn}

\section{Main Result}
\label{sec:MainResult}

\begin{table*}
\caption{Networks of interest and their capacity regions with and without security constraints.}
\label{table:MessaDescr}
\begin{center}
\begin{tabular}{ 
|>{\centering\arraybackslash}m{0.003\textwidth}|
|>{\centering\arraybackslash}m{0.2\textwidth}|
>{\centering\arraybackslash}m{0.2\textwidth}|
>{\centering\arraybackslash}m{0.2\textwidth}|  
>{\centering\arraybackslash}m{0.2\textwidth}|  
}
\hline
& & & & \\
\begin{turn}{90} {\bf{Network}}\end{turn}
&
\includegraphics[height=80px]{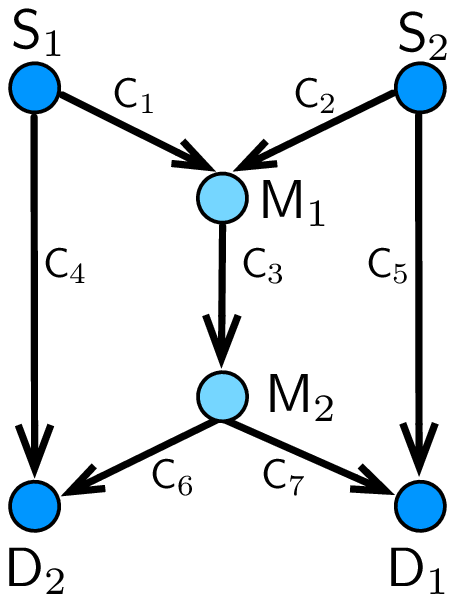} 
\captionof{figure}{Butterfly Network~1.}
\label{fig:butt1}
&
\includegraphics[height=80px]{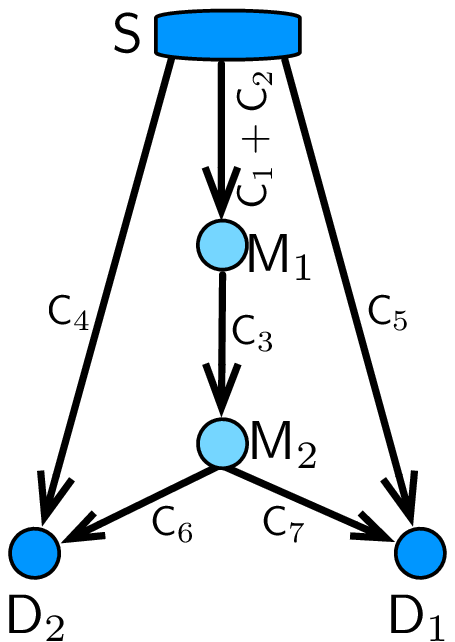} 
\captionof{figure}{Butterfly Network with Co-Located Sources.}
\label{fig:CS}
&
\includegraphics[height=80px]{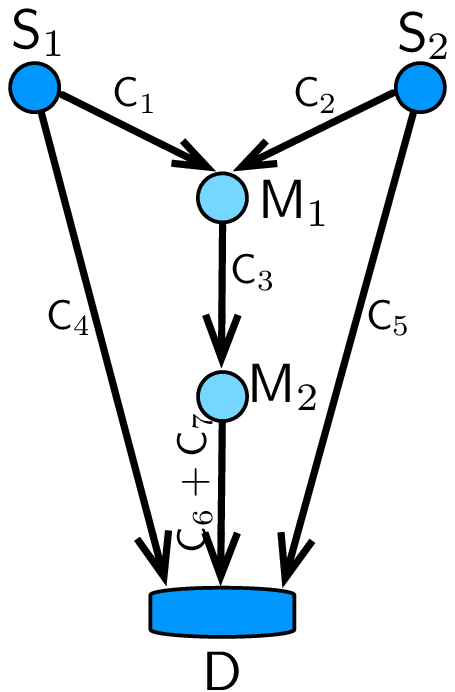} 
\captionof{figure}{Butterfly Network with Co-Located Sinks.}
\label{fig:CD}
&
\includegraphics[height=80px]{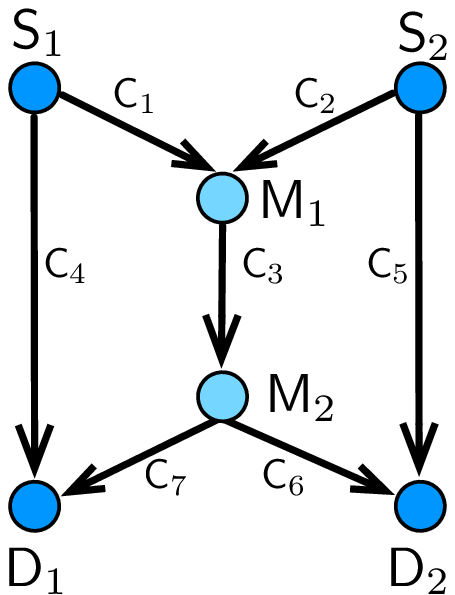} 
\captionof{figure}{Butterfly Network~2.}
\label{fig:butt2}
\\
\hline
\begin{turn}{90}{\bf{Without Security}}\end{turn}
&
{\begin{subequations}
\label{eq:but1cap}
\begin{align}
                         &R_1  \leq \min \left \{\mathsf{C}_1,\mathsf{C}_3,\mathsf{C}_7 \right \}, \label{eq:but1capc1} \\
         &R_2  \leq \min \left \{\mathsf{C}_2,\mathsf{C}_3,\mathsf{C}_6 \right \} \label{eq:but1capc2},
\\  &R_1 + R_2  \leq \mathsf{C}_3 + \mathsf{C}_4 \label{eq:but1capc3},
\\  &R_1 + R_2  \leq \mathsf{C}_3 + \mathsf{C}_5 \label{eq:but1capc4}.
            \end{align} \end{subequations}} 
&
{\begin{subequations}
\label{eq:CScap}
\begin{align}
	 &R_1  \!\leq \!\mathsf{C}_5 \!+\! \min \left \{ \mathsf{C}_1	\!+\!\mathsf{C}_2,\mathsf{C}_3,\mathsf{C}_7 \right \},
	\\ &R_2  \!\leq\! \mathsf{C}_4\!+\! \min \left \{ \mathsf{C}_1\!+\!\mathsf{C}_2,\mathsf{C}_3,\mathsf{C}_6 \right \},
	\\  &R_1 + R_2  \!\leq\! \mathsf{C}_4 \!+\! \mathsf{C}_5 \nonumber
\\& \quad +\! \min \left \{ \mathsf{C}_1\!+\!\mathsf{C}_2,\mathsf{C}_3,\mathsf{C}_6\!+\!\mathsf{C}_7 \right \}.
	\end{align}
\end{subequations}}
&
{\begin{subequations}
\label{eq:CDcap}
\begin{align}
	 &R_1 \!\leq\! \mathsf{C}_4 \!+\! \min \left \{ \mathsf{C}_1,\mathsf{C}_3,\mathsf{C}_6\!+\!\mathsf{C}_7\right \},
\\ &R_2  \!\leq\! \mathsf{C}_5\!+\! \min \left \{\mathsf{C}_2,\mathsf{C}_3,\mathsf{C}_6\!+\!\mathsf{C}_7 \right \},
\\ &R_1 + R_2   \!\leq\! \mathsf{C}_4 \!+\! \mathsf{C}_5\nonumber
\\& \quad +\! \min \left \{\mathsf{C}_1\!+\!\mathsf{C}_2,\mathsf{C}_3,\mathsf{C}_6\!+\!\mathsf{C}_7 \right \}.
	\end{align}
\end{subequations}}
&
{\begin{subequations}
\label{eq:Butt2cap}
\begin{align}
 &R_1  \leq \mathsf{C}_4 + \min \left \{\mathsf{C}_1,\mathsf{C}_3,\mathsf{C}_7 \right \},
	\\ &R_2 \leq \mathsf{C}_5+ \min \left \{ \mathsf{C}_2,\mathsf{C}_3,\mathsf{C}_6 \right \},
	\\  &R_1 + R_2 \leq \mathsf{C}_4 + \mathsf{C}_5+ \mathsf{C}_3.
	\end{align}
\end{subequations}}
\\
 \hline
\begin{turn}{90}{\bf{With Security}}\end{turn}
&
Secure communication is not possible. 
&
{\begin{subequations}
\label{eq:CScapSec}
	\begin{align}
& R_1 \leq \min \left \{\mathsf{C}_5,\mathsf{C}_1\!+\!\mathsf{C}_2,\mathsf{C}_3,\mathsf{C}_7 \right \}, \label{eq:bfcs_seq_ach_a}
\\	& R_2 \leq \min \left \{\mathsf{C}_4,\mathsf{C}_1\!+\!\mathsf{C}_2,\mathsf{C}_3,\mathsf{C}_6\right \}. \label{eq:bfcs_seq_ach_b}
	\end{align}
\end{subequations}}
&
{\begin{subequations}
\label{eq:CDcapSec}
	\begin{align}
		&R_1 \leq \min \left \{\mathsf{C}_1,\mathsf{C}_4 \right \}, \label{eq:bfcd_seq_ach_a}
		\\ &R_2  \leq \min \left \{ \mathsf{C}_2,\mathsf{C}_5\right \}, \label{eq:bfcd_seq_ach_b}
		\\  &R_1 + R_2 \leq \min \left \{ \mathsf{C}_3,\mathsf{C}_6+\mathsf{C}_7\right \}. \label{eq:bfcd_seq_ach_c}
	\end{align}
\end{subequations}}
&
{\begin{subequations}
\label{eq:Butt2capSec}
	\begin{align}
	 &R_1 \leq  \min \left \{ \mathsf{C}_4,\mathsf{C}_1,\mathsf{C}_3,\mathsf{C}_7\right \}, \label{eq:bf2_seq_ach_a}
	\\ &R_2 \leq  \min \left \{ \mathsf{C}_5,\mathsf{C}_2,\mathsf{C}_3,\mathsf{C}_6\right \}, \label{eq:bf2_seq_ach_b}
	\\  & R_1 + R_2  \leq \mathsf{C}_3. \label{eq:bf2_seq_ach_c}
	\end{align}
\end{subequations}}
\\
\hline   
\end{tabular}
\end{center}
\end{table*}

In this section we derive the capacity region with and without security constraints for the four $2$-unicast networks reported in Table~\ref{table:MessaDescr}.
In particular: 
(i) the network in Fig.~\ref{fig:butt1} is the classical butterfly network, which we refer to butterfly network 1;
(ii) the network in Fig.~\ref{fig:butt2} is a modified version of the classical butterfly network, which we refer to butterfly network 2;
(iii) the network in Fig.~\ref{fig:CS} is a particular case of the butterfly networks in Fig.~\ref{fig:butt1} and Fig.~\ref{fig:butt2} (with the role of $\mathsf{C}_4$ and $\mathsf{C}_5$ swapped) when the sources are co-located;
(iv) the network in Fig.~\ref{fig:CD} is a particular case of the butterfly networks in Fig.~\ref{fig:butt1} and Fig.~\ref{fig:butt2} when the destinations are co-located.
It is worth noting that when the two sources in Fig.~\ref{fig:butt1} are merged to get the network in Fig.~\ref{fig:CS}, the source is connected to $\mathsf{M}_1$ through two parallel edges of capacities $\mathsf{C}_1$ and $\mathsf{C}_2$. 
We replaced these two parallel edges by one edge of capacity $\mathsf{C}_1+\mathsf{C}_2$; 
while this operation is without loss of generality if there are no security constraints, it might be with loss of optimality under security considerations, since the eavesdropper can now wiretap an edge of capacity $\mathsf{C}_1+\mathsf{C}_2$, which is not possible in the case of parallel edges. 
However, our outer bounds (proved to be tight) are derived by considering the case where the eavesdropper never wiretaps the edge of capacity $\mathsf{C}_1+\mathsf{C}_2$. 
Hence, this operation is without loss of generality also under security considerations.
A similar argument holds for the case of co-located sinks, i.e., the two parallel edges of capacity $\mathsf{C}_6$ and $\mathsf{C}_7$ can be safely replaced by one edge of capacity $\mathsf{C}_6+\mathsf{C}_7$.
Table~\ref{table:MessaDescr} also reports, for each analyzed network, the capacity regions with and without security constraints, which are derived in the rest of this section.

\subsection{Butterfly Network 1}
We start by considering the butterfly network 1 in Fig.~\ref{fig:butt1} without security constraints. 
We notice that the rate region in~\eqref{eq:but1cap} is an outer bound on the capacity region of the butterfly network 1 in Fig.~\ref{fig:butt1} since: (i) the single rate constraints in~\eqref{eq:but1capc1} and~\eqref{eq:but1capc2} are cut-constraints (from the max-flow min-cut theorem) and (ii) the sum-rate constraints in~\eqref{eq:but1capc3} and~\eqref{eq:but1capc4} follow from the GNS outer bound in~\cite[Theorem 2]{kamath2011generalized}.
We now show that the rate region in~\eqref{eq:but1cap} is achievable.
In particular,
\begin{thm}
\label{thm:butterfly1}
For the butterfly network 1 in Fig.~\ref{fig:butt1}, the following rate region is achievable:
\begin{subequations}  
\label{eq:AchBut1}
	\begin{align}
	 R_1 & \leq \min \left (\mathsf{C}_1,\mathsf{C}_7 \right ), \label{eq:bf_ach_a}
	\\ R_2 &\leq \min \left (\mathsf{C}_2,\mathsf{C}_6 \right ), \label{eq:bf_ach_b}
	\\  R_1 + R_2 &\leq \mathsf{C}_3 + \min \left (R_2,\mathsf{C}_4, \mathsf{C}_5 \right ), \label{eq:bf_ach_c}
	\\ R_1 + R_2 &\leq \mathsf{C}_3 + \min \left (R_1,\mathsf{C}_4, \mathsf{C}_5 \right ). \label{eq:bf_ach_d}
	\end{align}
\end{subequations} 
\end{thm}
It is not difficult to see that the rate regions in~\eqref{eq:AchBut1} and~\eqref{eq:but1cap} are equivalent. 
Hence, the rate region in~\eqref{eq:but1cap} is the capacity region for the butterfly network 1 in Fig.~\ref{fig:butt1}.

\begin{IEEEproof}
Without loss of generality, we assume $R_1 \leq R_2$, i.e., the constraint in~\eqref{eq:bf_ach_c} is redundant.
Consider a rate pair $(R_1,R_2)$ satisfying the constraints in \eqref{eq:AchBut1}. 
The transmission scheme is as follows.
	\bit
	\item $\mathsf{S}_i, i \in [1:2]$ sends $R_i$ packets on edge $i$, which is possible thanks to the constraints in~\eqref{eq:bf_ach_a} and~\eqref{eq:bf_ach_b}.
	\item The intermediate node $\mathsf{M}_1$ is responsible of two operations: 
(i) it first merges $\min \left \{R_1,\mathsf{C}_4,\mathsf{C}_5 \right \}$ packets received from $\mathsf{S}_1$ with the same amount of packets received from $\mathsf{S}_2$, which is possible since we are assuming $R_1 \leq R_2$; after this first operation there are $\min \left \{R_1,\mathsf{C}_4,\mathsf{C}_5 \right \}$ mixed packets and $R_1 + R_2 - 2 \min \left \{R_1,\mathsf{C}_4,\mathsf{C}_5 \right \}$ uncoded packets (of which $R_i - \min \left \{R_1,\mathsf{C}_4,\mathsf{C}_5 \right \}$ were received from $\mathsf{S}_i, \ i \in [1:2]$); (ii) it then sends both these types of messages (i.e., coded and uncoded) on edge $3$, which is possible thanks to the constraint in~\eqref{eq:bf_ach_d}. 
\item The intermediate node $\mathsf{M}_2$ on edge $6$ (respectively, edge~$7$) sends: (i) $R_2 - \min \left \{R_1,\mathsf{C}_4,\mathsf{C}_5 \right \}$ (respectively, $R_1 - \min \left \{R_1,\mathsf{C}_4,\mathsf{C}_5 \right \}$) uncoded packets that were transmitted by $\mathsf{S}_2$ (respectively, $\mathsf{S}_1$) and (ii) all the $\min \left \{R_1,\mathsf{C}_4,\mathsf{C}_5 \right \}$ mixed packets received from $\mathsf{M}_1$; this operation is possible thanks to the constraint in~\eqref{eq:bf_ach_b}, i.e., $R_2 \leq \mathsf{C}_6$ (respectively, constraint in~\eqref{eq:bf_ach_a}, i.e., $R_1 \leq \mathsf{C}_7$).
\item Source $\mathsf{S}_1$ (respectively, $\mathsf{S}_2$) on edge $4$ (respectively, edge $5$) sends the $\min \left \{R_1,\mathsf{C}_4,\mathsf{C}_5 \right \}$ packets that were mixed at the intermediate node $\mathsf{M}_1$; notice that this operation is possible since $\min \left \{R_1,\mathsf{C}_4,\mathsf{C}_5 \right \} \leq \min \left \{\mathsf{C}_4,\mathsf{C}_5\right \}$.
	\item Destination $\mathsf{D}_1$ on edge $7$ receives $R_1 - \min \left (R_1,\mathsf{C}_4,\mathsf{C}_5 \right )$ uncoded packets of $\mathsf{S}_1$ and $\min \left (R_1,\mathsf{C}_4,\mathsf{C}_5 \right )$ packets of $\mathsf{S}_1$ mixed with same number of packets of $\mathsf{S}_2$, which are also received (uncoded) on edge $5$.
Thus, node $\mathsf{D}_1$ can recover the packets of $\mathsf{S}_1$ from the coded packets that it receives on edge $7$. 
Similarly, node $\mathsf{D}_2$ can successfully decode all the packets that were sent by source $\mathsf{S}_2$. 
\eit
\end{IEEEproof}

We now consider the butterfly network 1 in Fig.~\ref{fig:butt1} with security constraints. In particular,
\begin{thm}
\label{thm:butterfly1Secure}
For the butterfly network 1 in Fig.~\ref{fig:butt1}, secure communication is not possible.
\end{thm}

\begin{IEEEproof}
We consider block coding with block length $n$ and secret message rate $R_j, j \in [1:2]$.
With this, from the strong secrecy requirement
we obtain
\begin{align*}
n R_1 &\leq H \left ( W_1\right ) = I \left( W_1; Y_1^n \right) + H \left(W_1|Y_1^n \right) 
\\& < \epsilon_n + H \left(W_1|Y_1^n \right)
\\& =  I \left(W_1; Y_2^n,Y_5^n|Y_1^n \right) + H \left(W_1 | Y_1^n, Y_2^n, Y_5^n \right) + \epsilon_n
\\& \stackrel{{\rm{(a)}}}{ \leq } I \left(W_1; Y_2^n,Y_5^n|Y_1^n \right) + H \left(W_1 | Y_3^n, Y_5^n \right) + \epsilon_n
\end{align*}
\begin{align*}
& \stackrel{{\rm{(b)}}}{ \leq } I \left(W_1; Y_2^n,Y_5^n|Y_1^n \right) + H \left(W_1 | Y_7^n, Y_5^n \right) + \epsilon_n
\\& \stackrel{{\rm{(c)}}}{ \leq } I \left(W_1; Y_2^n,Y_5^n|Y_1^n \right) + n \epsilon_n + \epsilon_n
\\& = H \left( Y_2^n,Y_5^n|Y_1^n\right) \!-\! H \left(Y_2^n,Y_5^n|Y_1^n,W_1 \right)\!+\! n \epsilon_n \!+\! \epsilon_n
\\& \stackrel{{\rm{(d)}}}{ \leq } \!
H \left( Y_2^n,Y_5^n|Y_1^n\right) \!-\! H \left(Y_2^n,Y_5^n|Y_1^n,W_1, \Theta_1 \right)\!+\! n \epsilon_n \!+\! \epsilon_n
\\& \stackrel{{\rm{(e)}}}{ = }
H \left( Y_2^n,Y_5^n|Y_1^n\right) - H \left(Y_2^n,Y_5^n|W_1, \Theta_1 \right)+ n \epsilon_n + \epsilon_n
\\& \stackrel{{\rm{(f)}}}{ = }
H \left( Y_2^n,Y_5^n|Y_1^n\right) - H \left(Y_2^n,Y_5^n\right)+n \epsilon_n + \epsilon_n 
\\& \leq n \epsilon_n + \epsilon_n,
\end{align*}
where: 
(i) the inequality in $\rm{(a)}$ follows since $Y_3^n$ is a deterministic function of $\left( Y_1^n,Y_2^n\right)$ and because of the `conditioning reduces the entropy' principle;
(ii) the inequality in $\rm{(b)}$ follows since $Y_7^n$ is a deterministic function of $ Y_3^n$ and because of the `conditioning reduces the entropy' principle;
(iii) the inequality in $\rm{(c)}$ follows because of the decodability constraint;
(iv) the inequality in $\rm{(d)}$ is due to the `conditioning reduces the entropy' principle;
(v) the equality in $\rm{(e)}$ follows since $Y_1^n$ is a deterministic function of $\left( W_1, \Theta_1\right)$;
(vi) finally, the equality in $\rm{(f)}$ follows since $\left(Y_2^n,Y_5^n \right)$ is independent of $\left( W_1, \Theta_1\right)$.
By dividing both sides by $n$ and taking the limit for $n \rightarrow \infty$, we get $R_1 =0$. 
By following similar steps, one can derive $R_2 = 0$.
Hence, if do not have knowledge about the edge Eve is wiretapping, then a secure communication over the butterfly network 1 in Fig.~\ref{fig:butt1} is not possible.
\end{IEEEproof}

\subsection{Butterfly Network with Co-Located Sources}
We consider the butterfly network with co-located sources in Fig.~\ref{fig:CS} with no security constraints. 
We notice that the rate region in~\eqref{eq:CScap} is an outer bound on the capacity region of the network in Fig.~\ref{fig:CS} since all the rate constraints are cut-constraints (from the max-flow min-cut theorem).
We now show that the rate region in~\eqref{eq:CScap} is achievable.
In particular,
\begin{thm}
\label{thm:CS}
	For the butterfly network with co-located sources in Fig.~\ref{fig:CS}, the following rate region is achievable:	  
\begin{subequations}
\label{eq:AchButCS}
	\begin{align}
	 R_1 & \leq \mathsf{C}_7 + \min \left \{R_1,\mathsf{C}_5 \right \}, \label{eq:bfcs_ach_a}
	\\ R_2 & \leq \mathsf{C}_6 +  \min \left \{ R_2,\mathsf{C}_4 \right \}, \label{eq:bfcs_ach_b}
	\\ R_1 + R_2 & \leq \min \left \{ \mathsf{C}_1+\mathsf{C}_2,\mathsf{C}_3 \right \} + \min \left \{ R_2,\mathsf{C}_4 \right \} \nonumber
\\& \quad + \min \left \{R_1,\mathsf{C}_5 \right \}. \label{eq:bfcs_ach_c}
	\end{align}
\end{subequations}
\end{thm}
It is not difficult to see that the regions in~\eqref{eq:AchButCS} and~\eqref{eq:CScap} are equivalent. 
Hence, the rate region in~\eqref{eq:CScap} is the capacity region for the butterfly network with co-located sources in Fig.~\ref{fig:CS}.
\begin{IEEEproof}
Consider a rate pair $(R_1,R_2)$ satisfying the constraints in~\eqref{eq:AchButCS}. 
The transmission scheme is as follows.
	\bit
	\item The source sends $\min \left \{ R_1,\mathsf{C}_5 \right \}$ packets for $\mathsf{D}_1$ on edge $5$, which is possible since $\min \left \{ R_1,\mathsf{C}_5 \right \} \leq \mathsf{C}_5$. 
Similarly, the source sends $\min \left \{ R_2,\mathsf{C}_4 \right \}$ packets for $\mathsf{D}_2$ on edge $4$.
Moreover, on the link of capacity $\left(\mathsf{C}_1+\mathsf{C}_2 \right)$, the source sends $R_1 - \min \left \{ R_1,\mathsf{C}_5 \right \}$ packets for $\mathsf{D}_1$ and $R_2 - \min \left \{ R_2,\mathsf{C}_4 \right \}$ packets for $\mathsf{D}_2$. This operation is possible thanks to the constraint in~\eqref{eq:bfcs_ach_c}.
	\item The intermediate node $\mathsf{M}_1$ simply sends the $R_1 - \min \left \{ R_1,\mathsf{C}_5 \right \}$ packets for $\mathsf{D}_1$ and the $R_2 - \min \left \{ R_2,\mathsf{C}_4 \right \}$ packets for $\mathsf{D}_2$ on edge $3$, which is possible thanks to the constraint in~\eqref{eq:bfcs_ach_c}.
	\item The intermediate node $\mathsf{M}_2$ sends the $R_1 - \min \left \{ R_1,\mathsf{C}_5 \right \}$ packets for $\mathsf{D}_1$ (received from $\mathsf{M}_1$) on edge $7$, which is possible thanks to the constraint in~\eqref{eq:bfcs_ach_a}. Similarly, $\mathsf{M}_2$ sends the $R_2 - \min \left \{ R_2,\mathsf{C}_4 \right \}$ packets for $\mathsf{D}_2$ on edge $6$, which is possible thanks to the constraint in~\eqref{eq:bfcs_ach_b}.
	\item Node $\mathsf{D}_1$ (respectively, $\mathsf{D}_2$) successfully recovers a total of useful (i.e., those the source wished to explicitly communicate to $\mathsf{D}_1$) $R_1$ (respectively, $R_2$) uncoded packets. 
	\eit 
\end{IEEEproof}
We now consider the network in Fig.~\ref{fig:CS} with security constraints. In particular,
\begin{thm}
For the butterfly network with co-located sources in Fig.~\ref{fig:CS}, the secure capacity region is given by~\eqref{eq:CScapSec} in Table~\ref{table:MessaDescr}.
\end{thm}   
\begin{IEEEproof}
We here prove that the rate region in~\eqref{eq:CScapSec} in Table~\ref{table:MessaDescr} is achievable.
The proof that the rate region in~\eqref{eq:CScapSec} is also an outer bound on the secure capacity region of the butterfly network with co-located sources is reported in Appendix~\ref{app:OutBoundCS}.
Consider a secure rate pair $(R_1,R_2)$ satisfying the constraints in~\eqref{eq:CScapSec}. 
The transmission scheme is as follows.
	\bit
	\item The source sends $K=\max \left \{R_1,R_2 \right \}$ random packets on the edge of capacity $\mathsf{C}_1+\mathsf{C}_2$. 
These packets are used to generate a secret key. 
This operation is possible thanks to the constraints in~\eqref{eq:bfcs_seq_ach_a} and in~\eqref{eq:bfcs_seq_ach_b}.

\item The intermediate node $\mathsf{M}_1$ simply sends the $K$ random packets on edge $3$, which is possible thanks to the constraints in~\eqref{eq:bfcs_seq_ach_a} and in~\eqref{eq:bfcs_seq_ach_b}.

	\item The intermediate node $\mathsf{M}_2$ sends $R_1$ random packets (out the $K$ ones received from $\mathsf{M}_1$) on edge $7$, which we refer to as $K_1$. 
Similarly, out of the $K$ random packets received from $\mathsf{M}_1$, $\mathsf{M}_2$ sends $R_2$ random packets on edge $6$, which we refer to as $K_2$. 
These operations are possible thanks to the constraints in~\eqref{eq:bfcs_seq_ach_a} and in~\eqref{eq:bfcs_seq_ach_b}.

	\item The source sends $R_1$ message packets for $\mathsf{D}_1$ encrypted with the key $K_1$  on edge $5$ (possible because of~\eqref{eq:bfcs_seq_ach_a}). Similarly, it sends $R_2$ message packets for $\mathsf{D}_2$ encrypted with the key $K_2$ on edge $4$ (possible because of~\eqref{eq:bfcs_seq_ach_b}).

	\item Node $\mathsf{D}_1$ (respectively, $\mathsf{D}_2$) receives $R_1$ (respectively, $R_2$) encrypted useful packets from the source on edge $5$ (respectively, $4$). Hence, by using the key $K_1$ (respectively, $K_2$) received from $\mathsf{M}_2$, $\mathsf{D}_1$ (respectively, $\mathsf{D}_2$) successfully recovers $R_1$ (respectively, $R_2$) uncoded packets.
	\eit 
\end{IEEEproof}

\subsection{Butterfly Network with Co-Located Sinks}
We consider the butterfly network with co-located sinks as shown in Fig.~\ref{fig:CD} without security constraints. 
An outer bound on the capacity region of this network is the cut-set bound, which is given in~\eqref{eq:CDcap}.
We now design a transmission scheme that achieves the outer bound in~\eqref{eq:CDcap}. In particular,
\begin{thm}
	For the butterfly network with co-located sinks in Fig.~\ref{fig:CD}, the following rate region is achievable:	
\begin{subequations}
\label{eq:AchButCD}  
	\begin{align}
	 R_1 & \leq \mathsf{C}_1 + \min \left \{R_1,\mathsf{C}_4 \right \}, \label{eq:bfcd_ach_a}
	\\ R_2 & \leq \mathsf{C}_2 +  \min \left \{ R_2,\mathsf{C}_5 \right \}, \label{eq:bfcd_ach_b}
	\\ R_1 + R_2 &\leq \min \left \{ \mathsf{C}_3,\mathsf{C}_6+\mathsf{C}_7 \right \} + \min \left \{ R_2,\mathsf{C}_5 \right \} \nonumber
\\& \quad + \min \left \{ R_1,\mathsf{C}_4\right \}. \label{eq:bfcd_ach_c}
	\end{align}
\end{subequations}
\end{thm}
By straightforward manipulations, it is not difficult to see that the rate regions in~\eqref{eq:AchButCD} and~\eqref{eq:CDcap} are equivalent. 
Hence, the rate region in~\eqref{eq:CDcap} is the capacity region for the butterfly network with co-located sinks in Fig.~\ref{fig:CD}.
\begin{IEEEproof}
Consider a rate pair $(R_1,R_2)$ satisfying the constraints in \eqref{eq:AchButCD}. 
The transmission scheme is as follows.	\bit
	\item Source $\mathsf{S}_1$ (respectively, $\mathsf{S}_2$) sends $\min \left \{R_1,\mathsf{C}_4 \right \}$ (respectively, $\min \left \{R_2,\mathsf{C}_5 \right \}$) packets on edge $4$ (respectively, edge $5$).
Moreover, $\mathsf{S}_1$ (respectively, $\mathsf{S}_2$) sends $R_1 - \min \left \{R_1,\mathsf{C}_4\right \}$ (respectively, $R_2 - \min \left \{R_2,\mathsf{C}_5\right \}$) packets on edge $1$ (respectively, edge $2$). 
These operations are possible thanks to the constraints in~\eqref{eq:bfcd_ach_a} and in~\eqref{eq:bfcd_ach_b}.

	\item The intermediate nodes $\mathsf{M}_1$ and $\mathsf{M}_2$ simply send the $R_1 - \min \left \{R_1,\mathsf{C}_4 \right \}$ packets of $\mathsf{S}_1$ and the $R_2 - \min \left \{R_2,\mathsf{C}_5 \right \}$ packets of $\mathsf{S}_2$ on edge $3$ and on the edge of capacity $\mathsf{C}_6+\mathsf{C}_7$. This is possible thanks to the constraint in~\eqref{eq:bfcd_ach_c}.

	\item The destination successfully receives $R_1$ uncoded packets sent by $\mathsf{S}_1$ and $R_2$ uncoded packets sent by $\mathsf{S}_2$.
	\eit 
\end{IEEEproof}

We now consider the network in Fig.~\ref{fig:CD} with security constraints. In particular
\begin{thm}
For the butterfly network with co-located sinks in Fig.~\ref{fig:CD}, the secure capacity region is given in~\eqref{eq:CDcapSec} in Table~\ref{table:MessaDescr}.
\end{thm}   

\begin{IEEEproof}
We here prove that the rate region in~\eqref{eq:CDcapSec} in Table~\ref{table:MessaDescr} is achievable.
The proof of the converse is provided in Appendix~\ref{app:OutBoundCD}.
Consider a secure rate pair $(R_1,R_2)$ satisfying the constraints in~\eqref{eq:CDcapSec}. 
The transmission scheme is as follows.
\bit
	\item Source $\mathsf{S}_1$ (respectively, $\mathsf{S}_2$) sends $R_1$ (respectively, $R_2$) random packets on edge $1$ (respectively, $2$). 
These packets are used in the generation of the secret key and we refer to them to as $K_1$ (respectively, $K_2$). 
These operations are possible thanks to the constraints in~\eqref{eq:bfcd_seq_ach_a} and in~\eqref{eq:bfcd_seq_ach_b}.
Moreover, $\mathsf{S}_1$ (respectively, $\mathsf{S}_2$) sends $R_1$ (respectively, $R_2$) message packets encrypted with the key $K_1$ (respectively, $K_2$) on edge $4$ (respectively, edge $5$).
This is possible because of~\eqref{eq:bfcd_seq_ach_a} and~\eqref{eq:bfcd_seq_ach_b}.

	\item The intermediate nodes $\mathsf{M}_1$ and $\mathsf{M}_2$ simply send the $R_1+R_2$ random packets on edge $3$ and on the edge of capacity $\mathsf{C}_6+\mathsf{C}_7$, respectively. These operations are possible because of the constraints in~\eqref{eq:bfcd_seq_ach_c}.

	\item The destination receives $R_1$ encrypted packets of $\mathsf{S}_1$ on edge $4$ and $R_2$ encrypted packets of $\mathsf{S}_2$ on edge $5$. 
Moreover, it also receives the keys $K_1$ and $K_2$ on the edge of capacity $\mathsf{C}_6+\mathsf{C}_7$. 
Hence, by using the keys $K_1$ and $K_2$, it successfully recovers $R_1$ and $R_2$ uncoded packets of $\mathsf{S}_1$ and $\mathsf{S}_2$, respectively.
\eit 
\end{IEEEproof}

\subsection{Butterfly Network 2}
The last network we consider is the buttefly network 2 in Fig.~\ref{fig:butt2}, which differs from the buttefly network 1 in Fig.~\ref{fig:butt1} since each source is also directly connected to the corresponding destination. 
The rate region in~\eqref{eq:Butt2cap} is an outer bound on the capacity region of the butterfly network 2, where each constraint follows from the max-flow min-cut theorem.
We now show that the rate region in~\eqref{eq:Butt2cap} is achievable.
In particular,
\begin{thm}
	For the butterfly network 2 in Fig.~\ref{fig:butt2}, the following rate region is achievable:
\begin{subequations}  
\label{eq:AchBut2}  
	\begin{align}
	 R_1 &\leq \min \left \{ \mathsf{C}_1,\mathsf{C}_7 \right \} + \min  \left \{R_1,\mathsf{C}_4 \right \}, \label{eq:bf2_ach_a}
	\\ R_2 &\leq \min \left \{\mathsf{C}_2,\mathsf{C}_6 \right \} +  \min \left \{R_2,\mathsf{C}_5 \right \}, \label{eq:bf2_ach_b}
	\\ R_1 + R_2 & \leq \mathsf{C}_3 + \min \left \{R_2,\mathsf{C}_5 \right \} + \min \left \{R_1,\mathsf{C}_4 \right \}. \label{eq:bf2_ach_c}
	\end{align}
\end{subequations} 
\end{thm}
It is not difficult to see that the rate regions in~\eqref{eq:AchBut2} and~\eqref{eq:Butt2cap} are equivalent. 
Hence, the rate region in~\eqref{eq:Butt2cap} is the capacity region for the butterfly network 2 in Fig.~\ref{fig:butt2}.

\begin{IEEEproof}
Consider a rate pair $(R_1,R_2)$ satisfying the constraints in \eqref{eq:AchBut2}. 
The transmission scheme is as follows.
	\bit
	\item Source $\mathsf{S}_1$ (respectively, $\mathsf{S}_2$) sends $\min \left \{R_1,\mathsf{C}_4 \right \}$ (respectively, $\min \left \{R_2,\mathsf{C}_5 \right \}$) packets on edge $4$ (respectively, edge $5$).
Moreover, $\mathsf{S}_1$ (respectively, $\mathsf{S}_2$) sends $R_1 - \min \left \{R_1,\mathsf{C}_4 \right \}$ (respectively, $R_2 - \min \left \{R_2,\mathsf{C}_5 \right \}$) packets on edge $1$ (respectively, edge $2$). These operations are possible thanks to the constraints in~\eqref{eq:bf2_ach_a} and in~\eqref{eq:bf2_ach_b}.

	\item The intermediate node $\mathsf{M}_1$ sends $R_1 - \min \left \{R_1,\mathsf{C}_4 \right \}$ packets of $\mathsf{S}_1$ and $R_2 - \min \left \{R_2,\mathsf{C}_5 \right \}$ packets of $\mathsf{S}_2$ on edge $3$, which is possible because of the constraint in~\eqref{eq:bf2_ach_c}.

	\item The intermediate node $\mathsf{M}_2$ sends $R_1 - \min \left \{R_1,\mathsf{C}_4 \right \}$ packets of $\mathsf{S}_1$ on edge $7$. Similarly, $\mathsf{M}_2$ sends $R_2 - \min \left \{R_2,\mathsf{C}_5 \right \}$ packets of $\mathsf{S}_2$ on edge $6$. 
These operations are possible thanks to the constraints in~\eqref{eq:bf2_ach_a} and in~\eqref{eq:bf2_ach_b}.

	\item Node $\mathsf{D}_1$ (respectively, $\mathsf{D}_2$) receives $R_1$ (respectively, $R_2$) uncoded packets of $\mathsf{S}_1$ (respectively, $\mathsf{S}_2$).
	\eit 
\end{IEEEproof}
We now consider the network in Fig.~\ref{fig:butt2} with security constraints. In particular,
\begin{thm}
	For the butterfly network 2 in Fig.~\ref{fig:butt2}, the secure capacity region is given in~\eqref{eq:Butt2capSec} in Table~\ref{table:MessaDescr}.
\end{thm}   

\begin{IEEEproof}
We here prove that the rate region in~\eqref{eq:Butt2capSec} in Table~\ref{table:MessaDescr} is achievable.
The proof of the converse is provided in Appendix~\ref{app:OutBoundButt2}.
Consider a secure rate pair $(R_1,R_2)$ satisfying the constraints in~\eqref{eq:Butt2capSec}. 
The transmission scheme is as follows.
	\bit
	\item Source $\mathsf{S}_1$ (respectively, $\mathsf{S}_2$)
sends $R_1$ (respectively, $R_2$) random packets on edge $1$ (respectively, $2$). 
These packets are used in the secret key generation and we refer to them to as $K_1$ (respectively, $K_2$). 
Moreover, $\mathsf{S}_1$ (respectively, $\mathsf{S}_2$) sends $R_1$ (respectively, $R_2$) message packets encrypted with the key $K_1$ (respectively, $K_2$) on edge $4$ (respectively, $5$).
These operations are possible thanks to the constraints in~\eqref{eq:bf2_seq_ach_a} and in~\eqref{eq:bf2_seq_ach_b}.

	\item The intermediate node $\mathsf{M}_1$ simply sends the $R_1+R_2$ random packets ($K_1$ and $K_2$) on edge $3$. This operation is possible because of~\eqref{eq:bf2_seq_ach_c}.

	\item The intermediate node $\mathsf{M}_2$ sends the $R_1$ random packets of $\mathsf{S}_1$, i.e., $K_1$, on edge $7$.
Similarly, it sends the $R_2$ random packets of $\mathsf{S}_2$, i.e., $K_2$, on edge $6$. This is possible because of the constraints in~\eqref{eq:bf2_seq_ach_a} and~\eqref{eq:bf2_seq_ach_b}.

	\item The destination $\mathsf{D}_1$ (respectively, $\mathsf{D}_2$) receives  $R_1$ (respectively, $R_2$) encrypted message packets of $\mathsf{S}_1$ (respectively, $\mathsf{S}_2$). 
It also receives the key $K_1$ (respectively, $K_2$) on edge $7$ (respectively, $6$). 
Hence, by using $K_1$ (respectively, $K_2$), it successfully decodes $R_1$ (respectively, $R_2$) uncoded packets of $\mathsf{S}_1$ (respectively, $\mathsf{S}_2$).
	\eit 
\end{IEEEproof}

\section{Summary and Discussion}
\label{sec:summary}
In this paper we characterized the capacity of four two-unicast networks which are derived from the well-known butterfly network. 
In particular, we analyzed these networks with and without security constraints. 
Based on our analysis, we can draw the following conclusions.
\begin{enumerate}
\item There exist networks for which network coding operations are needed (for capacity characterization) in absence of security, but they do not provide any benefit under security constraints.
\item There are networks for which coding across sessions is not beneficial without security, but it becomes crucial with security considerations (see butterfly network with co-located sources).
\item Cooperation among sources and sinks increases the throughput both with and without security. 
We also observe that in case of security, if given an option of choosing between co-located sources or co-located destinations, the former brings higher throughput gains in case of uniform edge capacities. 
The results in this paper only consider the ultimate cooperation, i.e., we analyzed the case when the cooperation edge (between the two sources and the two sinks) is of infinite capacity (i.e., co-located nodes). Understanding how the rate advantages change with respect to the strength (finite capacity) of the cooperation link is an important open question, which is object of current investigation.
\end{enumerate}
The capacity characterization for a general multiple-unicast network is a long-standing open problem.
Several transmission schemes can be envisaged, one of which considers the butterfly network as a building block.
Hence, by means of the closed-form expression capacity result for the butterfly network (as well as for the networks derived from it) with general edge capacities, high-throughput achieving strategies for a general multiple-unicast network can be designed both with and without security constraints.

\appendices

\section{}
\label{app:OutBoundCS}
We consider block coding with block length $n$ and secret message rate $R_j, j \in [1:2]$.
We let $X_2^n$, $Y_2^n$ and $Z_2^n$ be the signal transmitted by the source on the edge of capacity $\mathsf{C}_1+\mathsf{C}_2$, the signal received by $\mathsf{M}_1$ and the signal received by the possible eavesdropper on the edge of capacity $\mathsf{C}_1+\mathsf{C}_2$, respectively.
With this we have
\begin{align}
n R_1 &\leq H \left ( W_1\right ) = I \left( W_1; Y_5^n \right) + H \left(W_1|Y_5^n \right)  \nonumber
\\& \stackrel{{\rm{(a)}}}{< } \epsilon_n + H \left(W_1|Y_5^n \right)\nonumber
\\& =  I \left(W_1; Y_7^n|Y_5^n \right) + H \left(W_1 | Y_5^n, Y_7^n\right) + \epsilon_n\nonumber
\\& \stackrel{{\rm{(b)}}}{ \leq }  I \left(W_1; Y_7^n|Y_5^n \right) + n \epsilon_n + \epsilon_n\nonumber
\\& \stackrel{{\rm{(c)}}}{\leq} H \left( Y_7^n|Y_5^n \right)+ n \epsilon_n + \epsilon_n\nonumber
\\& \stackrel{{\rm{(d)}}}{\leq} H \left( Y_7^n \right)+ n \epsilon_n + \epsilon_n \leq n \mathsf{C}_7+ n \epsilon_n + \epsilon_n,
\label{eq:constr1R1CS}
\end{align}
where: (i) the inequality in $\rm{(a)}$ follows because of the strong secrecy requirement;
(ii) the inequality in $\rm{(b)}$ follows because of the decodability constraint;
(iii) the inequality in $\rm{(c)}$ follows because the entropy of a discrete random variable is a non-negative quantity;
(iv) finally, the inequality in $\rm{(d)}$ is due to the `conditioning reduces the entropy' principle. 
By substituting the subscript $5$ with $7$ and vice versa in the above derivation, one can get the constraint $nR_1 \leq n \mathsf{C}_5+ n \epsilon_n + \epsilon_n$.
By means of similar steps, we obtain
\begin{align}
n R_1 & <
  I \left(W_1; Y_3^n|Y_5^n \right) + H \left(W_1 | Y_3^n, Y_5^n\right) + \epsilon_n \nonumber
\\ & \stackrel{{\rm{(a)}}}{\leq}
 I \left(W_1; Y_3^n|Y_5^n \right) + H \left(W_1 | Y_7^n, Y_5^n\right) + \epsilon_n \nonumber
\\ & \stackrel{{\rm{(b)}}}{\leq}
 I \left(W_1; Y_3^n|Y_5^n \right) + n \epsilon_n + \epsilon_n \nonumber
\\ & \stackrel{{\rm{(c)}}}{\leq}
 H \left( Y_3^n \right) + n \epsilon_n + \epsilon_n \leq n \mathsf{C}_3+ n \epsilon_n + \epsilon_n,
\label{eq:constr2R1CS}
\end{align}
where: 
(i) the inequality in $\rm{(a)}$ follows since $Y_7^n$ is a deterministic function of $ Y_3^n$ and because of the `conditioning reduces the entropy' principle;
(ii) the inequality in $\rm{(b)}$ follows because of the decodability constraint;
(iii) finally, the inequality in $\rm{(c)}$ follows because the entropy of a discrete random variable is non-negative and because of the `conditioning reduces the entropy' principle. 
By means of similar steps, we obtain
\begin{align}
n R_1 & <
  I \left(W_1; Y_2^n|Y_5^n \right) + H \left(W_1 | Y_2^n, Y_5^n\right) + \epsilon_n \nonumber
\\ & \stackrel{{\rm{(a)}}}{\leq}
 I \left(W_1; Y_2^n|Y_5^n \right) + H \left(W_1 | Y_3^n, Y_5^n\right) + \epsilon_n \nonumber
\\ & \stackrel{{\rm{(b)}}}{\leq}
 I \left(W_1; Y_2^n|Y_5^n \right)  + n \epsilon_n + \epsilon_n
\nonumber
\\ & \stackrel{{\rm{(c)}}}{\leq}
 H \left( Y_2^n \right) + n \epsilon_n + \epsilon_n \leq n \left(\mathsf{C}_1 +\mathsf{C}_2 \right)+ n \epsilon_n + \epsilon_n,
\label{eq:constr3R1CS}
\end{align}
where: 
(i) the inequality in $\rm{(a)}$ follows since $Y_3^n$ is a deterministic function of $ Y_2^n$ and because of the `conditioning reduces the entropy' principle;
(ii) the inequality in $\rm{(b)}$ follows from steps $\rm{(a)}$ and $\rm{(b)}$ in~\eqref{eq:constr2R1CS};
(iii) finally, the inequality in $\rm{(c)}$ follows because the entropy of a discrete random variable is a non-negative quantity and because of the `conditioning reduces the entropy' principle. 

By dividing both sides of the above inequalities by $n$ and by taking the limit for $n \rightarrow \infty$, we get the constraint in~\eqref{eq:bfcs_seq_ach_a}.
By following similar steps, one can derive the constraint in~\eqref{eq:bfcs_seq_ach_b}.

\section{}
\label{app:OutBoundCD}
We consider block coding with block length $n$ and secret message rate $R_j, j \in [1:2]$.
We let $X_6^n$, $Y_6^n$ and $Z_6^n$ be the signal transmitted by $\mathsf{M}_2$, the signals received by the destination and by the possible eavesdropper on the edge of capacity $\mathsf{C}_6+\mathsf{C}_7$, respectively.
With this we have
\begin{align}
n R_1 &\leq H \left ( W_1\right ) = I \left( W_1; Y_4^n \right) + H \left(W_1|Y_4^n \right)  \nonumber
\\& \stackrel{{\rm{(a)}}}{ < }  I \left(W_1; Y_5^n,Y_6^n|Y_4^n \right) + n \epsilon_n + \epsilon_n\nonumber
\\& \stackrel{{\rm{(b)}}}{ \leq }  I \left(W_1; Y_5^n,Y_3^n|Y_4^n \right) + n \epsilon_n + \epsilon_n\nonumber
\\& \stackrel{{\rm{(c)}}}{ \leq }  I \left(W_1; Y_1^n,Y_2^n,Y_5^n|Y_4^n \right) \!+\! n \epsilon_n \!+\! \epsilon_n\nonumber
\\& \stackrel{{\rm{(d)}}}{\leq} H \left( Y_1^n,Y_2^n,Y_5^n|Y_4^n \right) -H \left( Y_1^n,Y_2^n,Y_5^n|Y_4^n, W_1, \Theta_1 \right) \nonumber
\\& \quad + n \epsilon_n + \epsilon_n\nonumber
\\& \stackrel{{\rm{(e)}}}{=} H \left( Y_1^n,Y_2^n,Y_5^n|Y_4^n \right) -H \left( Y_1^n,Y_2^n,Y_5^n| W_1, \Theta_1 \right) \nonumber
\\& \quad + n \epsilon_n + \epsilon_n\nonumber
\\& \stackrel{{\rm{(f)}}}{=} H \left( Y_2^n,Y_5^n|Y_4^n \right)\!-\! H \left( Y_2^n,Y_5^n\right) \!+\! H \left( Y_1^n|Y_2^n,Y_4^n,Y_5^n\right) \nonumber
\\& \quad -H \left( Y_1^n| W_1, \Theta_1,Y_2^n,Y_5^n \right)+ n \epsilon_n + \epsilon_n \nonumber
\\& \stackrel{{\rm{(g)}}}{\leq} H \left( Y_1^n|Y_2^n,Y_4^n,Y_5^n\right) -H \left( Y_1^n| W_1, \Theta_1,Y_2^n,Y_5^n \right)\nonumber
\\& \quad + n \epsilon_n + \epsilon_n \nonumber
\\& \stackrel{{\rm{(h)}}}{\leq} H \left( Y_1^n\right) + n \epsilon_n + \epsilon_n \leq n \mathsf{C}_1 +n \epsilon_n + \epsilon_n,
\label{eq:constr1R1CD}
\end{align}
where: 
(i) the inequality in $\rm{(a)}$ follows because of the strong secrecy and the decodability constraints;
(ii) the inequality in $\rm{(b)}$ follows because of the `conditioning reduces the entropy' principle and since $Y_6^n$ is uniquely determined given $Y_3^n$; 
(iii) the inequality in $\rm{(c)}$ follows because of the `conditioning reduces the entropy' principle and since $Y_3^n$ is uniquely determined given $\left( Y_1^n, Y_2^n \right)$; 
(iv) the inequality in $\rm{(d)}$ is due to the `conditioning reduces the entropy' principle;
(v) the equality in $\rm{(e)}$ follows since $Y_4^n$ is a deterministic function of $\left( W_1, \Theta_1\right)$;
(vi) the equality in $\rm{(f)}$ follows since $\left( Y_2^n,Y_5^n\right)$ is independent of $\left(W_1, \Theta_1 \right)$;
(vii) the inequality in $\rm{(g)}$ is due to the `conditioning reduces the entropy' principle;
(viii) finally, the inequality in $\rm{(h)}$ follows since the entropy of a discrete random variable is non-negative and because of the `conditioning reduces the entropy' principle.
By means of similar steps as in~\eqref{eq:constr1R1CD} we obtain
\begin{align}
n R_1 &\leq H \left ( W_1\right ) = I \left( W_1; Y_1^n \right) + H \left(W_1|Y_1^n \right)  \nonumber
\\& < I \left(W_1; Y_2^n,Y_4^n,Y_5^n|Y_1^n \right) + n \epsilon_n + \epsilon_n\nonumber
\\& \leq H \left( Y_4^n|Y_1^n,Y_2^n,Y_5^n\right) -H \left( Y_4^n| W_1, \Theta_1,Y_2^n,Y_5^n \right)\nonumber
\\& \quad + n \epsilon_n + \epsilon_n \nonumber
\\& \leq H \left( Y_4^n\right) + n \epsilon_n + \epsilon_n \leq n \mathsf{C}_4 +n \epsilon_n + \epsilon_n.
\label{eq:constr2R1CD}
\end{align}
We now prove the outer bound in~\eqref{eq:bfcd_seq_ach_c}. In particular, we have
\begin{align}
&n \left(R_1+R_2 \right) \leq H \left ( W_1,W_2\right )  \nonumber
\\&=I \left( W_1,W_2; Y_4^n,Y_5^n \right) + H \left(W_1,W_2|Y_4^n,Y_5^n \right)  \nonumber
\\& \stackrel{{\rm{(a)}}}{<} I \left( W_1,W_2; Y_5^n|Y_4^n \right) + H \left(W_1,W_2|Y_4^n,Y_5^n \right) + \epsilon_n  \nonumber
\\& \stackrel{{\rm{(b)}}}{\leq} H \left( Y_5^n | Y_4^n\right)
- H \left(Y_5^n |  W_1, \Theta_1,W_2 \right) \nonumber
\\& \quad + H \left(W_1,W_2|Y_4^n,Y_5^n \right) + \epsilon_n  \nonumber
\\& \stackrel{{\rm{(c)}}}{\leq} I \left( W_2;Y_5^n | W_1, \Theta_1\right)
 + H \left(W_1,W_2|Y_4^n,Y_5^n \right) + \epsilon_n  \nonumber
\\& \stackrel{{\rm{(d)}}}{=} I \left( W_2;W_1, \Theta_1 | Y_5^n\right)+ I \left( W_2; Y_5^n\right) \nonumber
\\& \quad  + H \left(W_1,W_2|Y_4^n,Y_5^n \right) + \epsilon_n  \nonumber
\\& \stackrel{{\rm{(e)}}}{\leq} H\left( W_1, \Theta_1 | Y_5^n\right) -H\left( W_1, \Theta_1 | W_2, \Theta_2\right)  \nonumber
\\& \quad  + I \left( W_2; Y_5^n\right) + H \left(W_1,W_2|Y_4^n,Y_5^n \right) + \epsilon_n  \nonumber
\\& \stackrel{{\rm{(f)}}}{<}  H \left(W_1,W_2|Y_4^n,Y_5^n \right) + 2\epsilon_n  \nonumber
\\& \stackrel{{\rm{(g)}}}{\leq}  I \left(W_1,W_2; Y_6^n|Y_4^n,Y_5^n \right) + n \epsilon_n+ 2\epsilon_n  \nonumber
\\& \stackrel{{\rm{(h)}}}{\leq} H \left( Y_6^n\right) + n \epsilon_n + 2\epsilon_n \leq n \left(\mathsf{C}_6+\mathsf{C}_7 \right) +n \epsilon_n + \epsilon_n,
\label{eq:constr1R1R2CD}
\end{align}
where: 
(i) the inequality in $\rm{(a)}$ follows because of the strong secrecy constraint;
(ii) the inequality in $\rm{(b)}$ is due to the `conditioning reduces the entropy' principle and since $Y_4^n$ is a deterministic function of $\left( W_1, \Theta_1\right)$;
(iii) the inequality in $\rm{(c)}$ is due to the `conditioning reduces the entropy' principle and to the fact that $Y_5^n$ is independent of $\left( W_1, \Theta_1\right)$;
(iv) the equality in $\rm{(d)}$ follows since $W_2$ is independent of $\left( W_1, \Theta_1\right)$;
(v) the inequality in $\rm{(e)}$ follows because of the `conditioning reduces the entropy' principle and because $Y_5^n$ is uniquely determined given $\left( W_2, \Theta_2\right)$;
(vi) the inequality in $\rm{(f)}$ follows because of the strong secrecy constraint, because $\left( W_1, \Theta_1\right)$ is independent of $\left( W_2, \Theta_2\right)$ and because of the `conditioning reduces the entropy' principle;
(vii) the inequality in $\rm{(g)}$ follows from the decodability constraint;
(viii) finally, the inequality in $\rm{(h)}$ follows since the entropy of a discrete random variable is non-negative and because of the `conditioning reduces the entropy' principle.
Similarly,
\begin{align}
&n \left(R_1+R_2 \right) \stackrel{{\rm{(a)}}}{\leq}
I \left(W_1,W_2; Y_6^n|Y_4^n,Y_5^n \right) + n \epsilon_n+ 2\epsilon_n  \nonumber
\\& \stackrel{{\rm{(b)}}}{\leq}
I \left(W_1,W_2; Y_3^n|Y_4^n,Y_5^n \right) + n \epsilon_n+ 2\epsilon_n  \nonumber
\\& \stackrel{{\rm{(c)}}}{\leq} H \left( Y_3^n\right) + n \epsilon_n + 2\epsilon_n \leq n \mathsf{C}_3 +n \epsilon_n + \epsilon_n,
\label{eq:constr2R1R2CD}
\end{align}
where: 
(i) the inequality in $\rm{(a)}$ follows from the steps $\rm{(a)}$-$\rm{(g)}$ in~\eqref{eq:constr1R1R2CD};
(ii) the inequality in $\rm{(b)}$ follows because of the `conditioning reduces the entropy' principle and since $Y_6^n$ is uniquely determined given $Y_3^n$; 
(iii) finally, the inequality in $\rm{(c)}$ follows since the entropy of a discrete random variable is non-negative and because of the `conditioning reduces the entropy' principle.

By dividing both sides of the above inequalities by $n$ and by taking the limit for $n \rightarrow \infty$, we get the constraints in~\eqref{eq:bfcd_seq_ach_a} and in~\eqref{eq:bfcd_seq_ach_c}.
By following similar steps as in~\eqref{eq:constr1R1CD} and in~\eqref{eq:constr2R1CD}, one can derive the constraint on $R_2$ in~\eqref{eq:bfcd_seq_ach_b}.

\section{}
\label{app:OutBoundButt2}
We consider block coding with block length $n$ and secret message rate $R_j, j \in [1:2]$.
By following similar steps as in~\eqref{eq:constr1R1CS} with:
(i) the subscript $5$ replaced by $4$, it is straightforward to prove $n R_1 \leq n \mathsf{C}_7+ n \epsilon_n + \epsilon_n$;
(ii) the subscript $5$ replaced by $7$ and the subscript $7$ replaced by $4$, it is straightforward to prove $n R_1 \leq n \mathsf{C}_4+ n \epsilon_n + \epsilon_n$.
Similarly, by following the same steps as in~\eqref{eq:constr2R1CS} with the subscript $5$ replaced by $4$, one can easily prove $n R_1 \leq n \mathsf{C}_3+ n \epsilon_n + \epsilon_n$.
Moreover, we have
\begin{align}
n R_1 &\leq H \left ( W_1\right ) = I \left( W_1; Y_4^n \right) + H \left(W_1|Y_4^n \right)  \nonumber
\\ & \stackrel{{\rm{(a)}}}{<} I \left(W_1;Y_7^n|Y_4^n \right) +n \epsilon_n+ \epsilon_n  \nonumber
\\ & \stackrel{{\rm{(b)}}}{\leq} I \left(W_1;Y_3^n|Y_4^n \right) +n \epsilon_n+ \epsilon_n  \nonumber
\\ & \stackrel{{\rm{(c)}}}{\leq} I \left(W_1;Y_1^n,Y_2^n|Y_4^n \right) +n \epsilon_n+ \epsilon_n  \nonumber
\\ & \stackrel{{\rm{(d)}}}{\leq} H \left(Y_1^n,Y_2^n|Y_4^n \right)-H \left(Y_1^n,Y_2^n|Y_4^n,W_1, \Theta_1 \right) \nonumber
\\& \quad +n \epsilon_n+ \epsilon_n  \nonumber
\\ & \stackrel{{\rm{(e)}}}{\leq} H \left(Y_1^n,Y_2^n|Y_4^n \right)-H \left(Y_1^n,Y_2^n|W_1, \Theta_1 \right) \nonumber
\\& \quad +n \epsilon_n+ \epsilon_n  \nonumber
\\& \stackrel{{\rm{(f)}}}{=} H \left( Y_2^n|Y_4^n \right)- H \left( Y_2^n\right) + H \left( Y_1^n|Y_2^n,Y_4^n\right) \nonumber
\\& \quad -H \left( Y_1^n| W_1, \Theta_1,Y_2^n \right)+ n \epsilon_n + \epsilon_n \nonumber
\\& \stackrel{{\rm{(g)}}}{\leq}  H \left( Y_1^n|Y_2^n,Y_4^n\right)-H \left( Y_1^n| W_1, \Theta_1,Y_2^n \right)+ n \epsilon_n + \epsilon_n \nonumber
\\& \stackrel{{\rm{(h)}}}{\leq} H \left( Y_1^n\right) + n \epsilon_n + \epsilon_n \leq n \mathsf{C}_1 +n \epsilon_n + \epsilon_n,
\label{eq:constr4R1But2}
\end{align}
(i) the inequality in $\rm{(a)}$ follows because of the strong secrecy and the decodability constraints;
(ii) the inequality in $\rm{(b)}$ follows because of the `conditioning reduces the entropy' principle and since $Y_7^n$ is uniquely determined given $Y_3^n$; 
(iii) the inequality in $\rm{(c)}$ follows because of the `conditioning reduces the entropy' principle and since $Y_3^n$ is uniquely determined given $\left( Y_1^n,Y_2^n \right)$; 
(iv) the inequality in $\rm{(d)}$ is due to the `conditioning reduces the entropy' principle;
(v) the equality in $\rm{(e)}$ follows since $Y_4^n$ is a deterministic function of $\left( W_1, \Theta_1\right)$;
(vi) the equality in $\rm{(f)}$ follows since $Y_2^n$ is independent of $\left(W_1, \Theta_1 \right)$;
(vii) the inequality in $\rm{(g)}$ is due to the `conditioning reduces the entropy' principle;
(viii) finally, the inequality in $\rm{(h)}$ follows since the entropy of a discrete random variable is non-negative and because of the `conditioning reduces the entropy' principle.

We now prove the sum-rate outer bound in~\eqref{eq:bf2_seq_ach_c}. We have
\begin{align}
&n \left(R_1+R_2 \right) \stackrel{{\rm{(a)}}}{<} H \left(W_1,W_2|Y_4^n,Y_5^n \right) + 2\epsilon_n  \nonumber
\\ &\stackrel{{\rm{(b)}}}{\leq} I \left(W_1,W_2; Y_6^n,Y_7^n|Y_4^n,Y_5^n \right) + n \epsilon_n+ 2\epsilon_n  \nonumber
\\ &\stackrel{{\rm{(c)}}}{\leq} I \left(W_1,W_2; Y_3^n|Y_4^n,Y_5^n \right) + n \epsilon_n+ 2\epsilon_n  \nonumber
\\& \stackrel{{\rm{(d)}}}{\leq} H \left( Y_3^n\right) + n \epsilon_n + 2\epsilon_n \leq n \mathsf{C}_3 +n \epsilon_n +2 \epsilon_n,
\label{eq:constrR1R2But2}
\end{align}
where: 
(i) the inequality in $\rm{(a)}$ follows by similar steps as in $\rm{(a)}$-$\rm{(f)}$ in~\eqref{eq:constr1R1R2CD};
(ii) the inequality in $\rm{(b)}$ follows from the decodability constraint;
(iii) the inequality in $\rm{(c)}$ follows because of the `conditioning reduces the entropy' principle and since $\left(Y_6^n,Y_7^n\right)$ is uniquely determined given $Y_3^n$; 
(iv) finally, the inequality in $\rm{(d)}$ follows since the entropy of a discrete random variable is non-negative and because of the `conditioning reduces the entropy' principle.

By dividing both sides of the above inequalities by $n$ and by taking the limit for $n \rightarrow \infty$, we get the constraints in~\eqref{eq:bf2_seq_ach_a} and in~\eqref{eq:bf2_seq_ach_c}.
By following similar steps as done for deriving~\eqref{eq:bf2_seq_ach_a}, one can derive the constraint on $R_2$ in~\eqref{eq:bf2_seq_ach_b}.

\bibliographystyle{IEEEtran}
\bibliography{netcod2016}

\end{document}